\documentclass{article}
\usepackage[utf8]{inputenc}
\usepackage{graphicx}
\usepackage[portrait, margin=0.8in]{geometry}
\usepackage{hyperref}
\usepackage{authblk}
\usepackage{todonotes}
\usepackage{amsmath,amsthm,amssymb}
\usepackage{braket}
\usepackage{mathabx}
\usepackage{physics}
\usepackage[sorting = none, backend = bibtex, style=numeric-comp]{biblatex}

\title{Single Qubit Multi-Party Transmission Using Universal Symmetric Quantum Cloning}

\author[1]{Elijah Pelofske\thanks{Email: epelofske@lanl.gov}}
\affil[1]{Los Alamos National Laboratory}

\date{\vspace{-6ex}}

\begin{document}
\maketitle

\begin{abstract}
We consider the hypothetical quantum network case where Alice wishes to transmit one qubit of information (specifically a pure quantum state) to $M$ parties, where $M$ is some large number. The remote receivers locally perform single qubit quantum state tomography on the transmitted qubits in order to compute the quantum state within some error rate (dependent on the tomography technique and the number of transmitted qubits). We show that with the use of an intermediate optimal symmetric universal quantum cloning machine (between Alice and the remote receivers) as a repeater-type node in a hypothetical quantum network, Alice can send significantly fewer qubits compared to direct transmission of the message qubits to each of the $M$ remote receivers. This is possible due to two properties of quantum cloning. The first being that single qubit quantum clones retain the same Bloch angle as the initial quantum state. This means that if the mixed state of the quantum clone can be computed to high enough accuracy, the original pure quantum state can be inferred by extrapolating that vector to the surface of the Bloch sphere. The second property is that the state overlap of approximate quantum clones, with respect to the original pure quantum state, quickly converges (specifically for $1 \rightarrow M$ the limit of the fidelity as M goes to infinity is $\frac{2}{3}$). This means that Alice can prepare a constant number of qubits (which are then passed through the quantum cloning machine) in order to achieve a desired error rate, if $M$ is large enough. Combined, these two properties mean that for large $M$, Alice can prepare many orders of magnitude fewer qubits in order to achieve the same single qubit transmission accuracy compared to the naive direct qubit transmission approach. 

\end{abstract}

\section{Introduction}
\label{section:introduction}

\begin{figure}[h]
    \centering
    \includegraphics[width=1.0\textwidth]{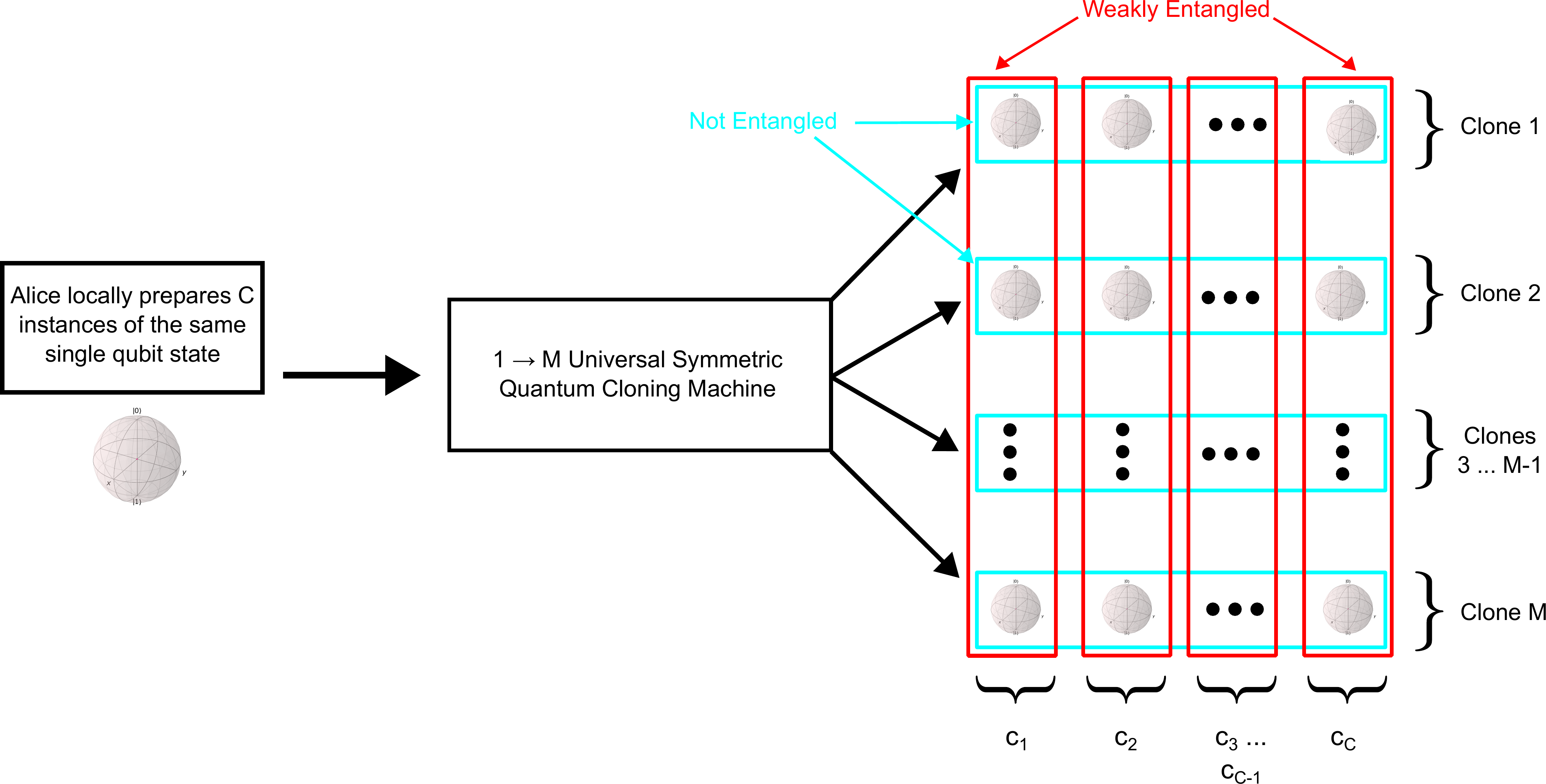}
    \caption{Qubit cloning network diagram, where the $1 \rightarrow M$ universal symmetric quantum cloning machine acts like a repeater for producing multiple approximate quantum clones. Each qubit that is fed into the quantum cloning machine results in $M$ approximate quantum clones that are weakly entangled (denoted in red). Alice can send $C$ independent instances of the same qubit through this cloning process, resulting in $C$ approximate clones (that are not entangled) being produced for each of the $M$ hypothetical receivers. To simplify the protocol, we assume that there are $M$ remote receivers, and the symmetric universal quantum cloning machine produces $M$ quantum clones from a single input qubit. Only the group of clone qubits produced by each execution of the symmetric universal quantum cloning machine will produce qubits which are weakly entangled. This hypothetical protocol is also assumed to be noiseless in the sense that this assumes no decoherence and no loss of qubits during transmission. }
    \label{fig:main_figure}
\end{figure}

Unknown quantum information can not, in general, be cloned - this is a fundamental property of quantum mechanics~\cite{wootters1982single, dieks1982communication}. However, approximate quantum cloning is possible~\cite{Bu_ek_1996}. In this study, we propose that universal, symmetric, optimal quantum cloning machines can be used in a repeater-type quantum network in order to transmit single qubits to a large number $M$ of remote receivers, where each remote receiver applies single qubit quantum state tomography on the received qubits in order to compute what the mixed density matrix state of the received quantum clones are, and then extrapolate what the original intended pure quantum state is (with some error rate). The proposed protocol is conceptually outlined in Figure \ref{fig:main_figure}. The use of qubits as the unit of transmitted information in this hypothetical protocol is motivated by the security given by the no cloning theorem itself and thus used in standard single qubit quantum key distribution~\cite{wootters1982single, dieks1982communication, Bennett_2014, Shor_2000, PhysRevA.72.012332, christandl2004genericsecurityproofquantum} -- in this case, specifically that measuring individual qubits that are being transmitted does not give sufficient information to fully reconstruct the original quantum state. Therefore, in this hypothetical scenario we also imagine that we wish to keep the information contained in the qubits used in this protocol secure as well. These single qubits transmitted in the proposed protocol have the same security inherent in quantum key distribution. 

Since the initial $1 \rightarrow 2$ quantum cloning was proposed~\cite{Bu_ek_1996}, it has been generalized to $N \rightarrow M$ quantum cloning~\cite{PhysRevLett.79.2153}. There are many variants of approximate quantum cloning \cite{Scarani_2005, fan2014quantum, PhysRevA.59.156, fiurasek2005highly, PhysRevA.67.022317, PhysRevA.62.012302, PhysRevA.72.052322, PhysRevLett.86.4942, hardy1999no, PhysRevLett.81.5003, PhysRevA.65.012304, PhysRevA.66.052111, PhysRevLett.87.247901, PhysRevA.72.042328, PhysRevA.60.136}, and there have been numerous experimental demonstrations of variants of quantum cloning \cite{PhysRevLett.88.187901, PhysRevLett.126.060503, PhysRevLett.106.180404, PhysRevLett.105.073602, bouchard2017high, PhysRevLett.94.040505, pelofske2023probing, Pelofske_2022_clone1, Pelofske_2022_clone2}. The primary characteristics that differentiate quantum cloning variants are as follows. \emph{Universal} means that the cloning process is input state independent, and non-universal means that the cloning process is state dependent. \emph{Symmetric} means that all generated clones have the same quality, whereas non-symmetric means that the generated clones can be of different quality (e.g., distinguishable). \emph{Optimal} quantum cloning means the process produces clones which are of the highest quality possible within the laws of quantum mechanics - this bound in terms of quantum fidelity is given by eq.~\eqref{eq:theoretical-fidelity}: 

\begin{equation}
    F_{N \rightarrow M} = \frac{ MN + M + N }{ M(N+2) }.
    \label{eq:theoretical-fidelity}
\end{equation}

Whereas typically in quantum networks one of the central goals is to share entanglement using entanglement swapping \cite{PhysRevLett.128.150502, PhysRevLett.119.170502, PhysRevLett.101.080403, PhysRevA.90.032306, PhysRevA.81.052329, PhysRevA.77.062301}, there are a number of proposed algorithms that encode information into individual qubits, usually in the context of quantum machine learning \cite{PerezSalinas2020datareuploading, thumwanit2021trainable, ambainis2009quantum, Pe_a_Tapia_2023}. Therefore, it is conceivable that in a hypothetical future large scale quantum network, one may wish to transmit a single qubit state to a large number of remote parties. We can imagine two potential cases where the proposed method could be used. The first is where the preparation of the single qubit state requires a non-insignificant amount of compute time, and therefore Alice wishes to reduce the total number of preparations (in particular to offload the compute time onto the quantum cloning process). The second is where a a quantum cloning machine already has a direct networked connection to the intended recipients, and therefore it is easier for Alice to send qubits through the quantum cloning node in order to distribute the quantum information.

Accurately measuring what quantum state has been prepared, particularly on hardware experiments, is of considerable interest for quantum information processing. Quantum state tomography of a single qubit~\cite{Schmied_2016} in particular is the simplest of these types of tasks, since there is not an exponential overhead that comes with larger system sizes. Therefore, in this study we consider strictly the case where the recipient(s) measure the state of single qubits using quantum state tomography. There are many methods for performing full quantum state tomography, but in this case we use Pauli basis state tomography in all simulations. Geometrically, the factor by which the Bloch vector of an input message state in the Bloch sphere representation is shrunk by when copied by an optimal universal symmetric quantum cloning process is given by 

\begin{equation}
    \eta (N, M) = \frac{ N }{ M } \frac{M+2}{N+2}, 
    \label{eq:shrinking_factor}
\end{equation}

see refs.~\cite{PhysRevLett.81.2598, PhysRevA.57.2368, Scarani_2005} (for a qubit, e.g., $d=2$).

\begin{figure}[h]
    \centering
    \includegraphics[width=0.49\textwidth]{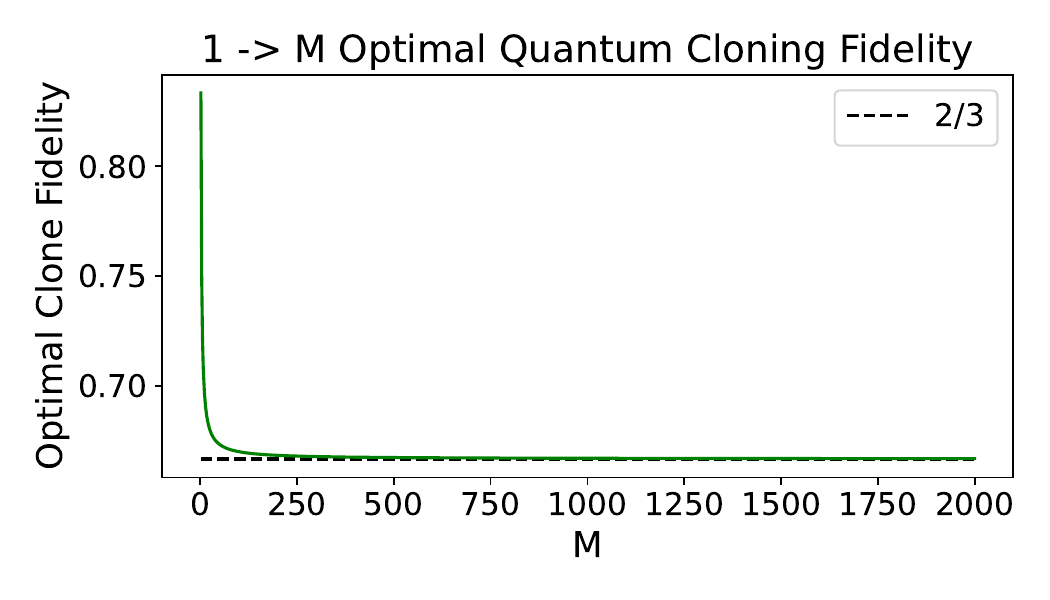}
    \caption{Ideal quantum clone fidelity (eq. \eqref{eq:theoretical-fidelity}) for $1 \rightarrow M$ universal cloning as a function of $M$, for $M=2 \dots 2000$. }
    \label{fig:ideal_fidelity}
\end{figure}

For universal symmetric quantum cloning machines, there is a known optimal bound on the best quantum state fidelity that can be achieved for single qubit clones. This bound is shown in eq.~\eqref{eq:theoretical-fidelity}. The primary motivation of this proposed single qubit distribution methodology is that $\lim_{M \to\infty} F_{1 \rightarrow M} = \frac{2}{3}$. This fidelity convergence is plotted in Figure \ref{fig:ideal_fidelity}. This means that the single qubit clone quality is asymptotic, which in particular means that there are \emph{not} diminishing returns as $M$ gets extremely large, in terms of error measures such as the state overlap between any one of the approximate quantum clones and the original message qubit. In this context, the property of quantum cloning machines that gives this asymptotic fidelity quality is that \emph{asymptotic quantum cloning} (where the number of clones tends to infinity) is the same as direct quantum state estimation, as shown in refs. \cite{PhysRevLett.97.030402, https://doi.org/10.4230/lipics.tqc.2013.220}. Quantum state estimation is the general task of learning information about a quantum state using measurements~\cite{paris2004quantum}, and quantum state tomography is a specific type of quantum state estimation.

Refs.~\cite{9782866, s23187891} studied similar ideas of using quantum cloning for qubit transmission, but limited the number of clones up to $4$, and did not use single qubit quantum state tomography (QST) or geometric extrapolation to determine the intended pure quantum state. Refs.~\cite{wang2019duplicating, wang2018filling} studied a similar ideas of using many qubit copies to transmit information, but only examined making repeated quantum copies of classical bits, and did not extend the number of clones to large $M$.

\section{Methods}
\label{section:methods}

In order to quantify how close the post processed quantum clones are to the original pure quantum state, we measure the distance along the surface of the Bloch sphere between the extrapolated single qubit clone vector (since universal quantum cloning only shrinks the Bloch vector by a factor given in eq.~\eqref{eq:shrinking_factor}, and does not change the angle of the vector) and the pure quantum state vector at the point they intersect the surface of the Bloch sphere. In other words, this error measure quantifies the geodesic distance along the Bloch sphere between two Bloch vectors. This measure is $0$ when the two states are exactly the same and is at a maximum of $\pi$ when the vectors are pointing in opposite directions. The distance between the two points, along the surface of the Bloch sphere, is computed by $\arccos (\rho_x \cdot \rho_{\text{clone}_x} + \rho_y \cdot \rho_{\text{clone}_y} + \rho_z \cdot \rho_{\text{clone}_z})$, where $\rho_x, \rho_y, \rho_z$ are the x, y, z coordinates for the intersection point on the surface of the Bloch sphere for the pure quantum state, and $\rho_{\text{clone}_x}, \rho_{\text{clone}_y}, \rho_{\text{clone}_z}$ are the x, y, z coordinates for the intersection point of the extrapolated vector of the single qubit clone density matrix. Note that due to numerical precision error, occasionally this term will be outside of $[-1, 1]$ making it undefined for $\arccos$, in which case the computed value is set to $-1$ or $1$, respectively. The implementation of the numerical simulation protocol is specified with the goal of showing how relatively easy to implement this protocol is conceptually -- the difficulty of the proposed protocol is strictly due to the quantum information processing and quantum networking that would be required to actually physically implement it. 

The extrapolation procedure we employ is to compute the coordinates of the mixed state density matrix within the Bloch sphere, then compute the intersection of that Bloch vector with the surface of the Bloch sphere. The extrapolated Bloch vector intersection point with the Bloch sphere can be computed by solving for the positive value of $t$ in $(t\cdot x)^2 + (t\cdot y)^2 + (t\cdot z)^2 - 1 = 0$, where $x$, $y$, and $z$ are the coordinates of the mixed state Bloch vector. Then the extrapolated $x, y, z$ coordinates are given by $x \cdot t$, $y \cdot t$ and $z \cdot t$. This is easily computed over many numerical simulations using sympy~\cite{10.7717/peerj-cs.103}. The error of the extrapolated point can then be measured by the distance along the surface of the Bloch sphere, as described above, or can be measured by converting the coordinates into a density matrix form and the state overlap with the message state can be measured. The second metric we will use to quantify the error rate of the single qubit state tomography (both for the standard approach and the proposed quantum cloning approach) is the well established quantum fidelity~\cite{jozsa1994fidelity, müller2023simplified} metric, which measures the state overlap between two density matrices, and is given in eq.~\eqref{eq:fidelity-general}. Fidelity of $0$ means there is no state overlap, and a fidelity of $1$ means that the two states are exactly overlapping. The error rate will be reported as one minus the fidelity, e.g., the infidelity. The quantum fidelity measure, which is an overlap measure between two density matrices, is given as

\begin{equation}
    F(\rho_1, \rho_2) = Tr[ \sqrt{ \sqrt{\rho_1}\rho_2 \sqrt{\rho_1} } ]^2. 
    \label{eq:fidelity-general}
\end{equation}

The density matrix reconstruction was performed using a slightly modified version of Qiskit Ignis~\cite{Qiskit}, with the least squares parameter optimization performed using the Python 3 package cvxopt~\cite{diamond2016cvxpy, agrawal2018rewriting}, which uses convex optimization \cite{agrawal2019dgp, agrawal2020dqcp, agrawal2019differentiable, agrawal2020differentiating} in order to perform maximum likelihood estimation \cite{Smolin_2012}. Conversion formulas between density matrix representations and Bloch sphere coordinates are given in Appendix~\ref{section:appendix_density_matrix_formulas}. All simulations done in this study assume no decoherence of any of the quantum states during transmission (or any other part of the protocol), and assume in general ideal conditions - for example we assume no qubit loss during transmission either. The only noise we consider is shot noise (e.g., finite sampling effect). 

Quantum fidelity is a standard quantum state overlap measure, and therefore is most likely to be interpretable. However, the distance along the surface of the Bloch sphere provides a geometric intuition which is very compatible with the notion of single-qubit quantum cloning, and therefore we report both error measures in this study.

\subsection{Clone Emulation}
\label{section:methods_clone_emulation}
Since single qubit clones of universal, symmetric, quantum cloning machines have well defined properties, namely that the clones of quantum state correspond geometrically to the vector of the original state being shrunk by a factor $\eta$, given by eq. \eqref{eq:shrinking_factor}.

This means that for the purposes of analyzing un-entangled sequences of single qubit clones (e.g., the un-entangled sequences described in Figure~\ref{fig:main_figure}), one can emulate the relevant quantum cloning procedure by constructing a density matrix that describes the mixed state of the original pure quantum state, shrunk by $\eta$. Importantly, this involves only a single qubit, and thus classical simulations of the state are extremely fast, and this can be executed for any $M$. We refer to this procedure as \emph{clone emulation} in order to make it clear that this is not producing a full set of $M$ quantum clones (which themselves form an entangled system), and is a limited but extremely useful tool for analyzing single qubit quantum cloning in this proposed protocol. It should emphasized however that this cloning emulation is only a numerical simulation method that allows us to examine the expected error rates and overhead of this hypothetical networking protocol. If one were to implement this protocol, the full universal quantum cloning machine would need to implement a very large and complex unitary, thus generating $M$ weakly entangled clones. Exact statevector simulation of such a quantum cloning unitary, for very large $M$, is not feasible because of the exponentially growing compute time and memory cost of full statevector simulation of quantum states. 

Note that in this study we will denote $M$ simultaneously as the number of approximate quantum clones generated by the $1 \rightarrow M$ cloning process, and the number of remote clone receivers illustrated in Figure~\ref{fig:main_figure}. In other words, each of the $M$ clones that is produced given an input of a single original pure quantum single-qubit state, is sent to exactly one (independent) remote receiver, who then performs local operations and measurements. Additionally, for all simulations the same original message qubit is repeatedly transmitted to all $M$ parties so that a baseline of error rate metrics can be measured as other parameters are changed.

For all numerical simulations, a single qubit state is used since the cloning process is universal (state independent). For the clone emulation procedure, $10,000$ instances (for each number of shots) are executed, instead of a full $M$ instances being executed. This only works because the cloning is symmetric (i.e., all quantum clones are identical), and because the remote receiver measurements are always from separate groups of clones (cyan boxes in Figure~\ref{fig:main_figure}), and therefore not entangled. Note that if we were interested in computing the full quantum state of a group of clones (e.g., the red boxes in Figure~\ref{fig:main_figure}), the computational cost would be immense for large $M$ - in particular this would require at least $M$ qubits with no or very little errors, and a significant number of quantum gates to be executed (see refs.~\cite{pelofske2023probing, Pelofske_2022_clone1, Pelofske_2022_clone2} for detailed circuit descriptions of quantum telecloning circuits, as an example). Alternatively, if the form of the cloning state density matrix was known then it could be computed and sampled from - however this would require a matrix that has dimensions $2^M$ by $2^M$. For the scenario in Figure~\ref{fig:main_figure}, it is not required that these full state simulations be performed in order to assess the transmission error rates that each remote receiver would experience. In particular, for universal and symmetric quantum cloning the clone quality is significantly simplified -- single qubit state tomography on approximate quantum clones can show what the resulting transmission accuracy would be (after single qubit quantum state tomography and state reconstruction). 

In order to verify that the clone emulation works as expected, we perform simulations that compare classical simulations of quantum cloning circuits against the single qubit emulation. Specifically, this comparison is made using parallel single qubit quantum state tomography run using full statevector simulations of quantum cloning circuits in order to compute (ground-truth) clone quality. The quantum cloning circuits that we use for simulations of this protocol are a variant of quantum cloning, known as quantum telecloning, that are universal, symmetric, and optimal. The construction of these telecloning circuits is described in refs.~\cite{pelofske2023probing, Pelofske_2022_clone1, Pelofske_2022_clone2}. Appendix~\ref{section:appendix_quantum_telecloning_circuit} shows an example of one of these circuits, described in the form of an explicit compiled quantum circuit diagram.

\begin{figure}[h!]
    \centering
    \includegraphics[width=0.49\textwidth]{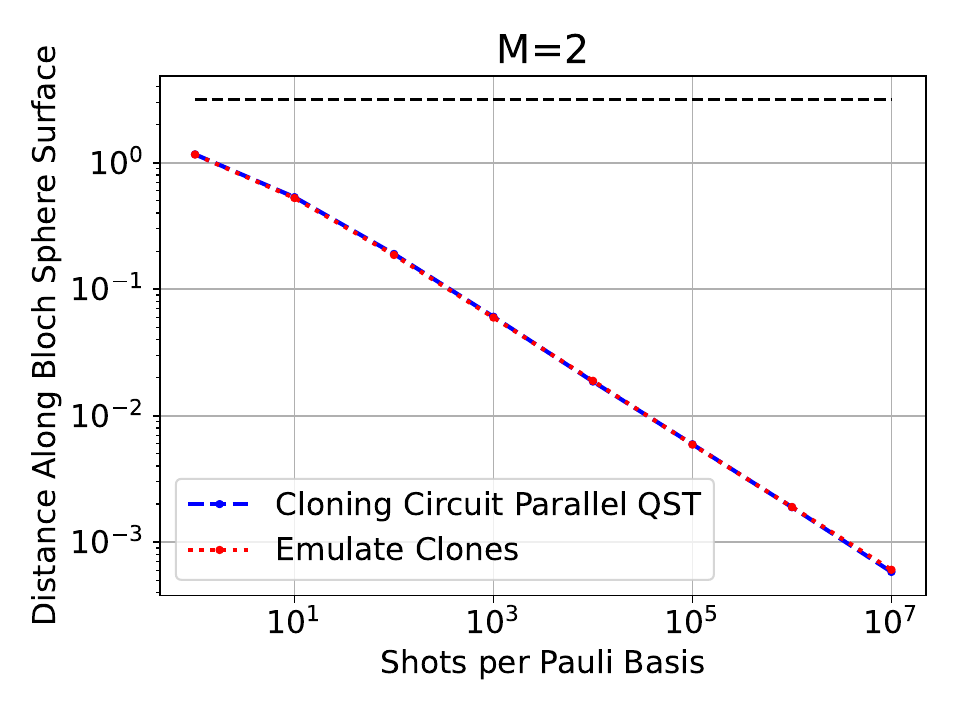}
    \includegraphics[width=0.49\textwidth]{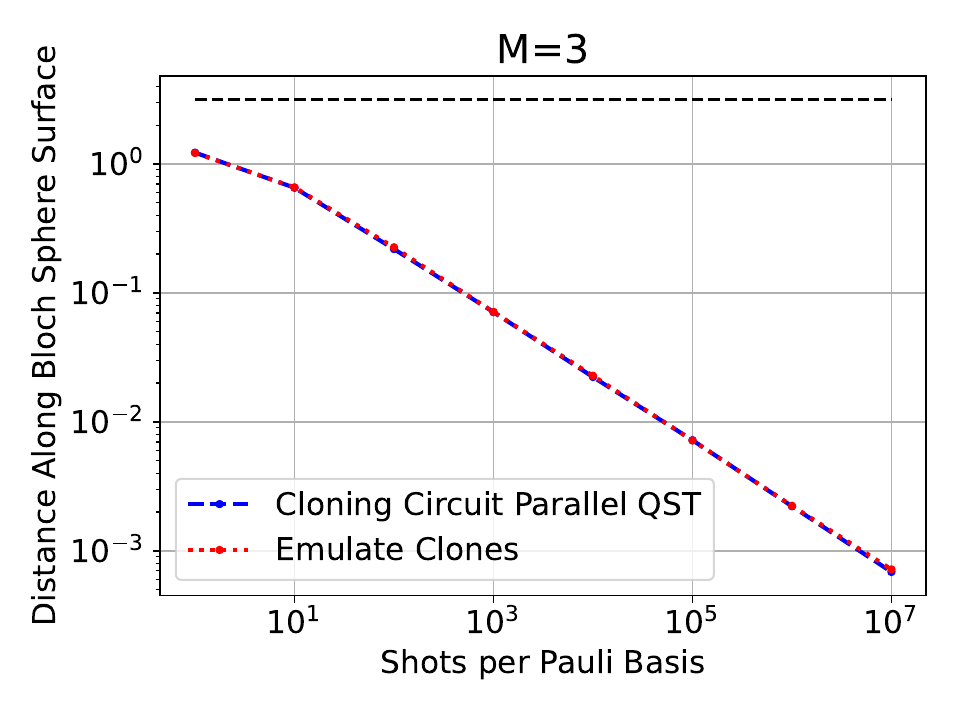}
    \caption{Comparison of single qubit clone emulation against full quantum circuit parallel single qubit state tomography for $M=2$ and $M=3$. For each point, the mean geodesic distance from the intended point on the Bloch sphere is plotted, averaged over $1,000$ separate simulations ($3 \cdot M \cdot 1000$ measurements are taken in total for both procedures for each point, although because the cloning is symmetric these properties hold for any of the $M$ clones). The high agreement shown in these plots demonstrates that the clone emulation trick matches full (parallel measured) quantum state tomography of a small (example) universal quantum cloning machine unitary, implemented as a quantum circuit. Data is plotted on a log-log scale axis. }
    \label{fig:compare_parallel_QST_to_emulate}
\end{figure}

\section{Results}
\label{section:results}

Figure~\ref{fig:compare_parallel_QST_to_emulate} verifies that the clone emulation procedure produces identical results, up to shot noise, compared to the full parallel single qubit state tomography procedure. The full parallel single qubit state tomography procedure was performed using the quantum telecloning circuits from refs.~\cite{pelofske2023probing, Pelofske_2022_clone1, Pelofske_2022_clone2}. 

\begin{figure}[h!]
    \centering
    \includegraphics[width=0.84\textwidth]{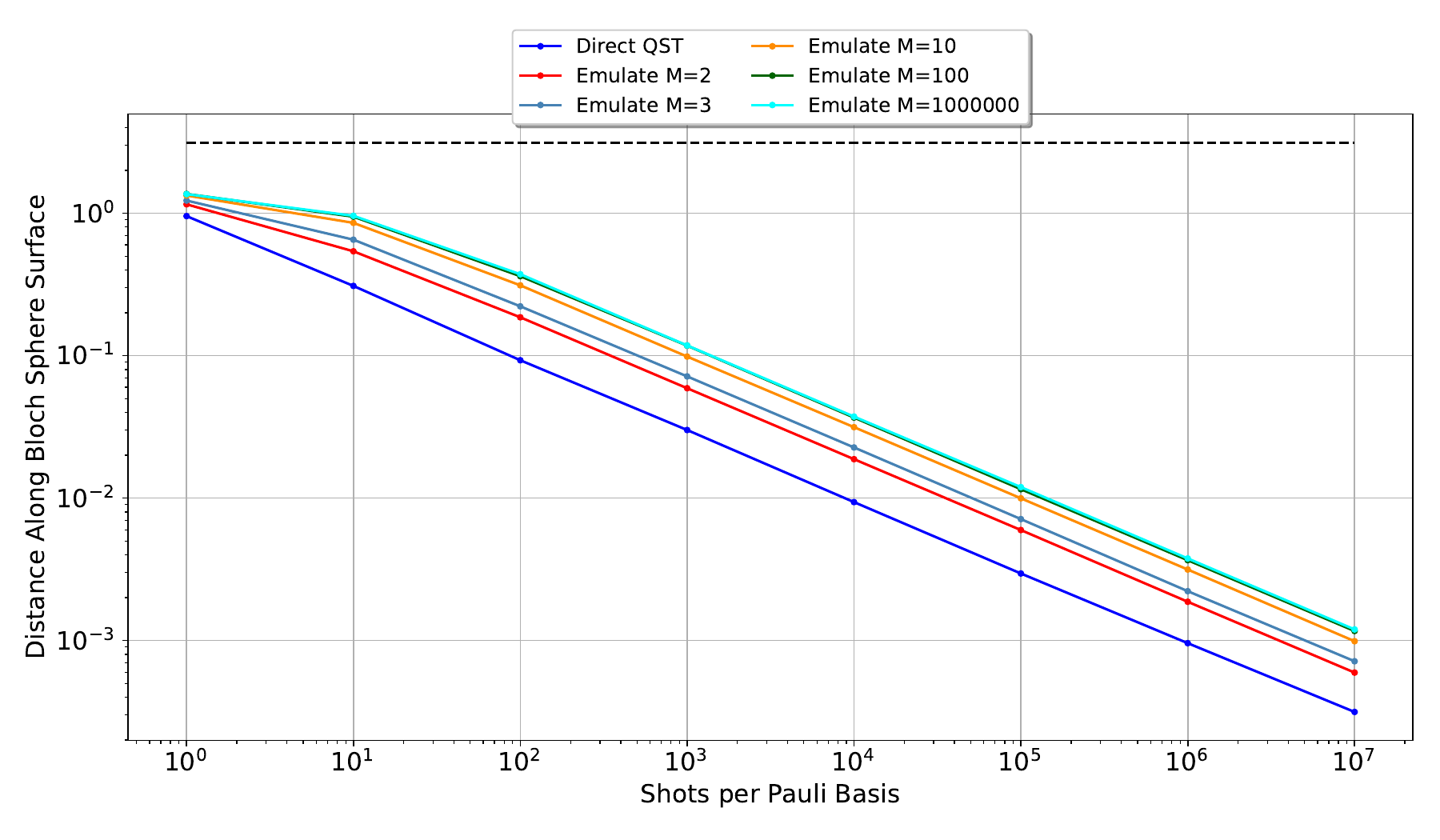}
    \includegraphics[width=0.84\textwidth]{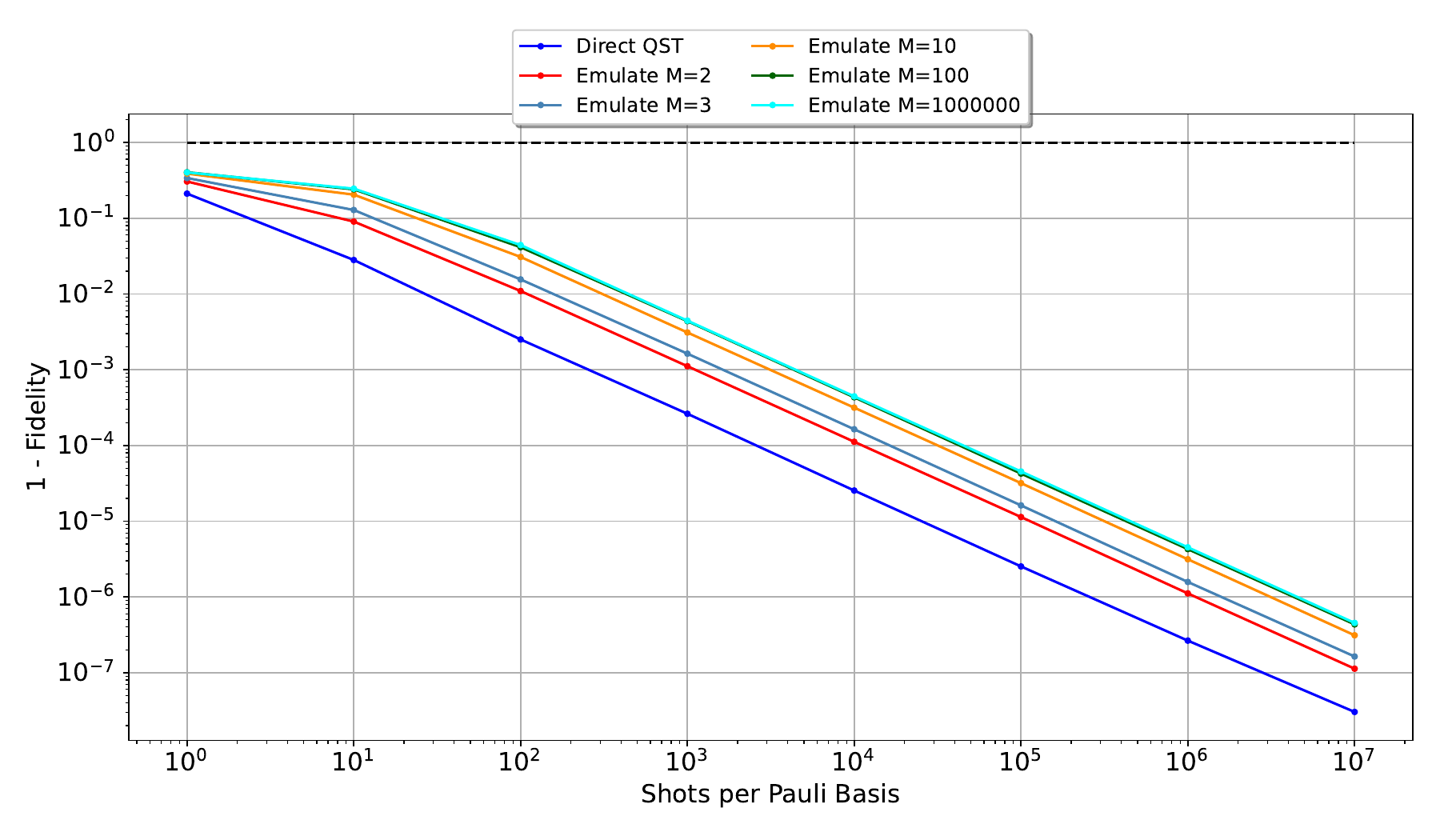}
    \caption{Comparison of direct quantum state tomography and quantum state tomography on single qubit emulated clones, using the metric of mean distance along the Bloch sphere surface (top) and mean infidelity, which is 1 minus the standard state overlap quantum fidelity measure (bottom). Both metrics quantify the error of the transmission protocol, when the recipient(s) measure using quantum state tomography, therefore for both error rate measures closer to $0$ corresponds to a lower error rate. Due to the convergence of the state overlap between the single qubit clones and the original state (eq~\eqref{eq:theoretical-fidelity}), for example the error rate of $M=100$, and $M=10^6$ have converged to nearly identical values, resulting in those lines being visually nearly identical. This figure shows that ss expected when transmitting single qubits the direct qubit transmission (labeled as Direct QST) gives the lowest error rate. The maximum error rates for both metrics are plotted as black horizontal dashed lines. Data is plotted on a log-log scale axis. }
    \label{fig:compare_direct_QST_to_clone}
\end{figure}

\begin{figure}[h!]
    \centering
    \includegraphics[width=0.49\textwidth]{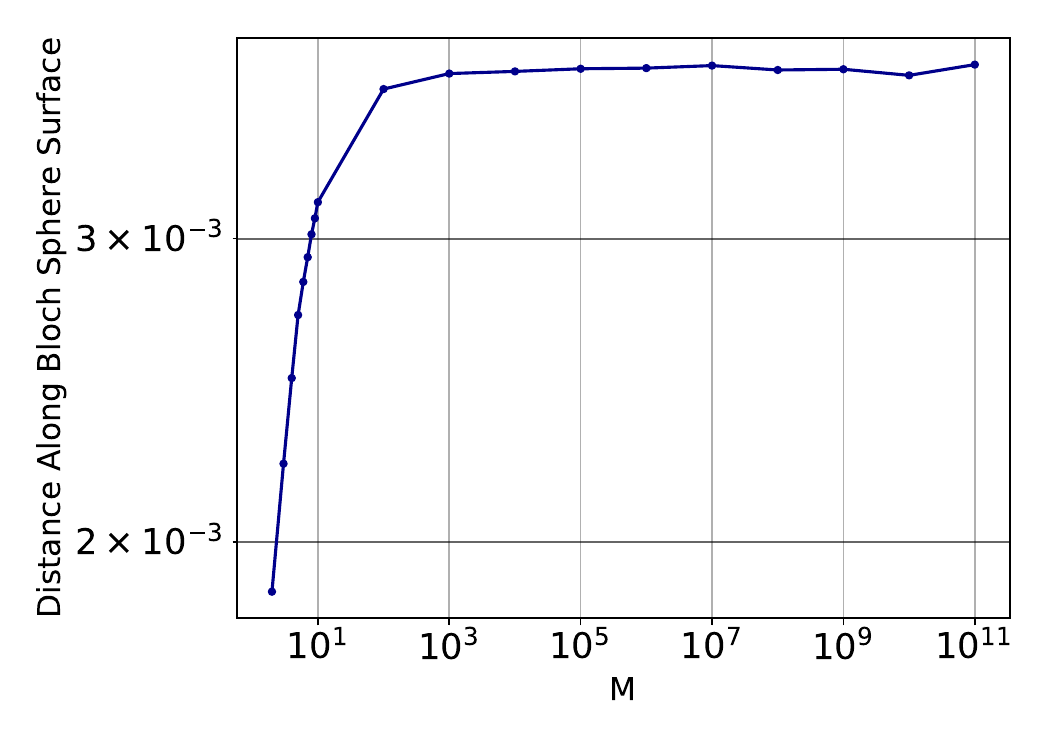}
    \includegraphics[width=0.49\textwidth]{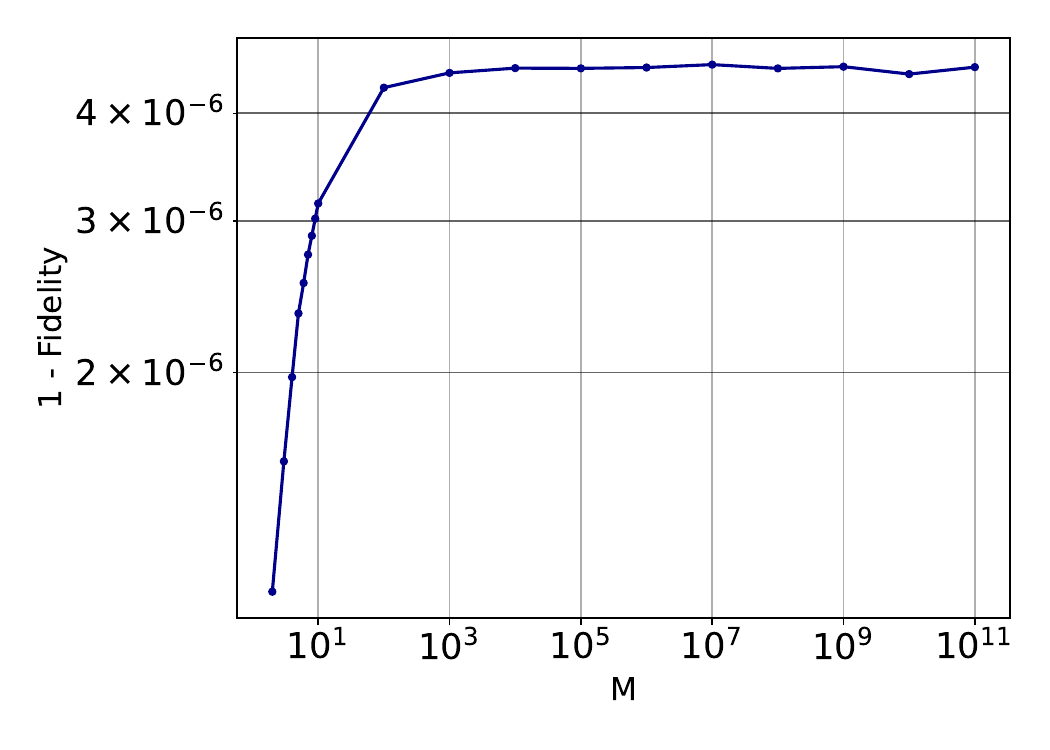}
    \caption{Average error rate (y-axis) as a function of $M$ when using $1 \rightarrow M$ quantum cloning. Both the error rate metrics of quantum infidelity (right) and geodesic distance along the surface of the Bloch sphere between the two vectors (left) are both shown. A fixed number of shots, $10^6$ (per Pauli basis, meaning $3\cdot10^6$ in total), is used to estimate these error rates. These plots show a clear convergence of error rates while $M$ is increasing. Log-log scale axis. }
    \label{fig:increasing_M_error_rate}
\end{figure}

Figure~\ref{fig:compare_direct_QST_to_clone} shows a comparison between direct single qubit state tomography and extrapolated quantum state reconstruction from the approximate quantum clones. Direct QST refers to the procedure where Alice transmits the qubit to each of the $M$ recipients, repeated many times (plotted on the x-axis of Figure~\ref{fig:compare_direct_QST_to_clone}). The metrics used for this comparison are i) distance along the surface of the Bloch sphere (closer to $0$ means lower error rate), and ii) $1 - F$ (closer to $0$ means lower error rate). As expected, because $\lim_{M \to\infty} F_{1 \rightarrow M} = \frac{2}{3}$, there is an asymptote of the error rate of the reconstructed quantum states as $M$ gets large; $M=100$ and $M=10^6$ are effectively visually indistinguishable, which makes sense because all of the generated clones at this scale are approaching the same state overlap with the pure quantum state. The data Figure~\ref{fig:compare_direct_QST_to_clone} represents the mean error rate from $1,000$ separate executions of quantum state tomography, for each point on the x-axis. The emulated quantum clones are generated in batched simulations of $10,000$ per parameter. 

Figure~\ref{fig:compare_direct_QST_to_clone} shows that the proposed protocol (see Figure~\ref{fig:main_figure}) provides a significant improvement over direct single qubit transmission when $M$ is large (but when considering only single qubits and in particular single clones of the original qubit, as expected, direct QST gives the lowest error rate). The tradeoff that this protocol proposes is to sacrifice some error rate in the small $M$ regime in order to make the total number of transmitted copies of that original qubit very large, and at very large $M$ we get a constant error rate as shown by Figure~\ref{fig:increasing_M_error_rate}. Note that because of the asymptotic clone quality, $M$ can be any large finite number. Figure~\ref{fig:increasing_M_error_rate} shows this asymptotic clone quality as $M$ becomes extremely large, for both error metrics, for a fixed number of shots. This shows that, for sufficiently large $M$, Alice can use a constant number of qubits (for a desired error rate) in order to transmit those qubits, via a symmetric universal quantum cloning machine, to $M$ remote receivers. Figure~\ref{fig:compare_direct_QST_to_clone} shows that there is approximately an order of magnitude separation between the limiting behavior of the quantum cloning and the direct qubit transmission. In order to quantify this, we can measure the breakeven point between these two transmission methods.

\begin{figure}[h!]
    \centering
    \includegraphics[width=0.72\textwidth]{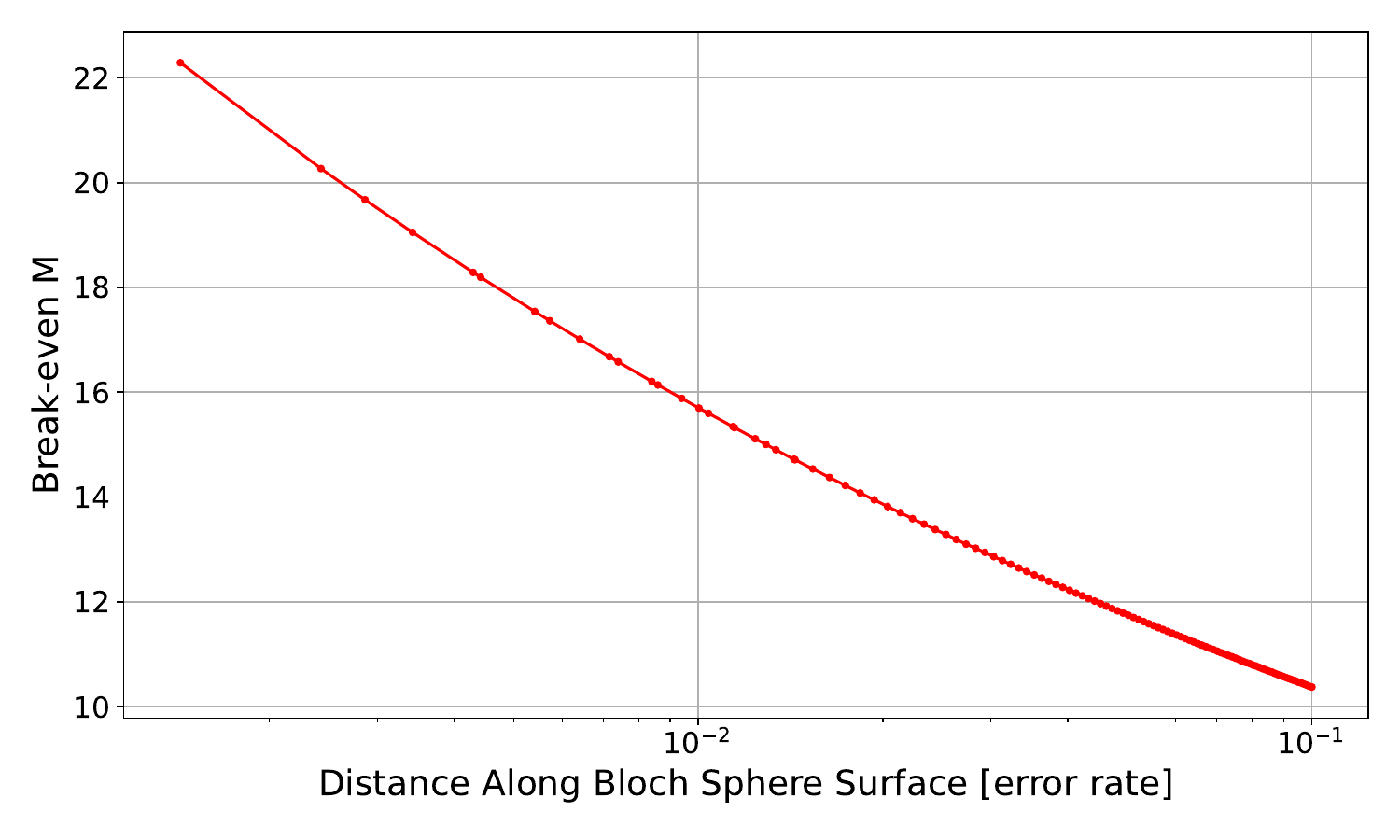}
    \includegraphics[width=0.72\textwidth]{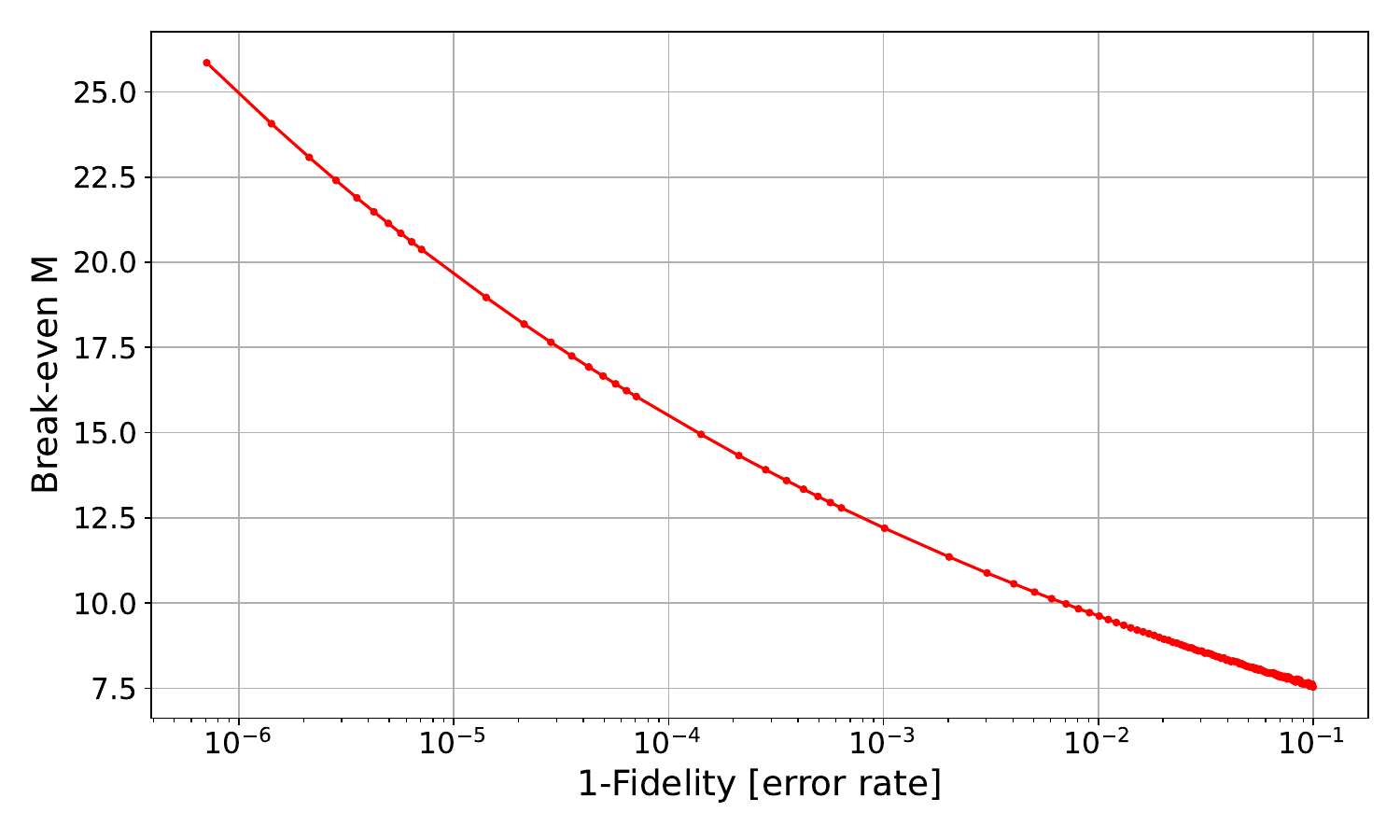}
    \caption{Average breakeven point in terms of $M$ (y-axis), as a function of the two error rate metrics (x-axis); distance along the surface of the Bloch sphere (top) and quantum infidelity (bottom). The breakeven point we define as the point where the two single qubit transmission methods require Alice to prepare the same number of message qubits in order for the transmission error rate to be equivalent (for all $M$ remote receivers) when the single qubit states are being measured using Pauli basis quantum state tomography. x-axis is log scale. }
    \label{fig:breakeven}
\end{figure}

The breakeven point is defined as the point where the two methods (the direct single qubit transmission and the proposed quantum cloning protocol) require the same number of message qubits to be prepared by Alice in order for the receiver error rate (for all $M$ receivers) to be the same. Figure~\ref{fig:breakeven} plots this breakeven point as a function of error rate. The breakeven point for this range of error rates was computed using an interpolation between the datapoints in Figure~\ref{fig:compare_direct_QST_to_clone}. 

For a desired error rate, if $M$ is greater than the break-even point in Figure \ref{fig:breakeven}, Alice must produce a constant number of message qubits (given by the error-dependent curve, approximated to high accuracy by the $M=100000$ curve in Figure \ref{fig:compare_direct_QST_to_clone}) to be passed through the quantum cloning machine (Figure \ref{fig:main_figure}). In order for the direct single qubit transmission, without quantum cloning, to achieve the same error rate and throughput (e.g., sending the qubits to each of the $M$ remote receivers), Alice would need to prepare $M$ qubits, per Pauli basis (in this case 3), for each sample shown by the x-axis of the blue \textit{Direct QST} curve in Figure~\ref{fig:compare_direct_QST_to_clone}. This shows a clear and significant scaling advantage of using quantum cloning to perform this type of single qubit transmission compared to direct qubit transmission without cloning. 

Figure~\ref{fig:breakeven} contains three key observations. First is that the scaling of the breakeven point as a function of $M$ is increasing - as the transmission error rate decreases, $M$ increases. This increase of $M$ is not significant (it reaches $25$ in Figure \ref{fig:breakeven}), but a thorough analysis of this scaling for substantially smaller error rates is an interesting question for future work. Second, the breakeven point is similar but not identical for the two error metrics. Third, the breakeven point scaling, at least for the tested error rate range, is incredibly favorable to the proposed quantum cloning protocol. Namely that in order for Alice to see a reduction in the number of local qubit preparations, $M$ (the number of remote receivers) must be greater than $25$. And importantly, as shown by the convergence of the transmission error rates for large $M$ in Figure~\ref{fig:increasing_M_error_rate}, the number of qubits that Alice must prepare locally for any desired error rate (or better) is then constant for any larger value of $M$.

\section{Discussion and Conclusion}
\label{section:discussion}
This study has proposed a hypothetical use for large scale universal symmetric quantum cloning machine in the context of transmitting a single qubit from a local sender Alice to a large number of remote receivers $M$, where the remote receivers receive multiple transmitted qubits and perform single qubit quantum state tomography in order to reconstruct the intended quantum state. We have shown that the use of a large scale quantum cloning node, in a hypothetical quantum network, can provide significant reduction in the total number of qubits that Alice must prepare locally. This occurs when the transmitted qubits can be sent through a universal symmetric quantum cloning machine, and when $M$ is sufficiently large (shown by Figure~\ref{fig:breakeven}). This method works due to two properties of optimal symmetric universal quantum cloning;

\begin{enumerate}
    \item Single qubit clones retain the same Bloch angle as the parent clone, in the Bloch sphere representation; the loss of fidelity corresponds to a shrinking of the Bloch vector, making the clone a mixed quantum state when the original quantum state was pure. This means that given a sufficient number of samples of an approximate quantum clone, generated from the same input quantum state, the original quantum state can be extrapolated by extending the computed mixed state Bloch vector to the surface of the Bloch sphere. 
    \item The state overlap, e.g., fidelity, of the clones generated by $1 \rightarrow M$ quantum cloning converges quickly to $\frac{2}{3}$ in the limit as $M$ goes to infinity. This means that there is a limit on the reduction of clone quality as $M$ becomes large. 
\end{enumerate}

The primary logical next question is to what extent more general (universal, symmetric) $N \rightarrow M$ quantum cloning processes could be used in this type of quantum information distribution protocol, since typically entangled states contain more interesting algorithmic information that one may wish to distribute over a quantum network. Specifically, the most important question is how the extrapolation procedure can be extended to entangled multi-qubit systems. It would be of interest to determine to what extend entangled quantum states can be accounted for in this type of quantum cloning extrapolation protocol. This certainly should be possible, however the clear Bloch sphere geometric argument used in this paper would not easily extend to $N \geq 2$ - most likely an extrapolation method based purely on the density matrices would be necessary.

The emphasis of this study is on showing a proof of principle for a specific type of quantum networking protocol. Importantly, this type of protocol is far from currently feasible on real quantum computers or quantum networks -- and this study does not examine real world aspects that would certainly be relevant, such as timing. Quantifying how real world decoherence error rates, both for the transmission and for the preparation of the quantum cloning circuit, impact the reconstruction error is another important future question. Noise, in particular decoherence, could both shrink the length of the Bloch vector as well as bias the Bloch vector angle. Nevertheless, the implementation shown here uses the real world characterization protocol of quantum state tomography, thus showing that this could be implemented algorithmically if there existed hardware that could carry out these operations (namely that it would need to be possible to get good low error rate state reconstruction despite a very large numbers of clones being generated in total). 

Although not investigated in this paper, this proposed protocol could certainly also be applied to message qubits which are mixed states, but this would cause the cloning fidelity to be even lower than it is when the message qubit is a pure quantum state, thus necessitating more copies to passed through the quantum cloning machine in order to get reasonably low error rate single qubit state tomography for the receivers.

\section{Acknowledgments}
\label{section:acknowledgments}
This work was supported by the U.S. Department of Energy through the Los Alamos National Laboratory. Los Alamos National Laboratory is operated by Triad National Security, LLC, for the National Nuclear Security Administration of U.S. Department of Energy (Contract No. 89233218CNA000001). Research presented in this article was supported by the NNSA's Advanced Simulation and Computing Beyond Moore's Law Program at Los Alamos National Laboratory. This research used resources provided by the Darwin testbed at Los Alamos National Laboratory (LANL) which is funded by the Computational Systems and Software Environments subprogram of LANL's Advanced Simulation and Computing program (NNSA/DOE). LANL report number LA-UR-23-31425.

\appendix
\section{Density Matrix and Bloch Sphere Conversion Formulas}
\label{section:appendix_density_matrix_formulas}
In order to convert from x, y, z coordinates (of the head of the Bloch vector representing the quantum state) into a density matrix, the following matrix form can be used ($i$ denotes the imaginary component):

$$
\begin{bmatrix}
0.5 + 0.5z & 0.5x-0.5y i  \\
0.5x + 0.5y i & 0.5-0.5z  \\
\end{bmatrix}
$$

\noindent
In order to convert from the following density matrix into Bloch sphere coordinates, 

$$
\begin{bmatrix}
a & b  \\
c & d  \\
\end{bmatrix}
$$

\noindent
The $x$ coordinate is $\Re(c+b)$, the $y$ coordinate is $\Im{c-b}$ and the $z$ coordinate is $\Re(d-a)$.

\section{Distribution of Error Rates for the Quantum Cloning Transmission}
\label{section:appendix_distribution_of_error_rates}

\begin{figure}[h!]
    \centering
    \includegraphics[width=0.49\textwidth]{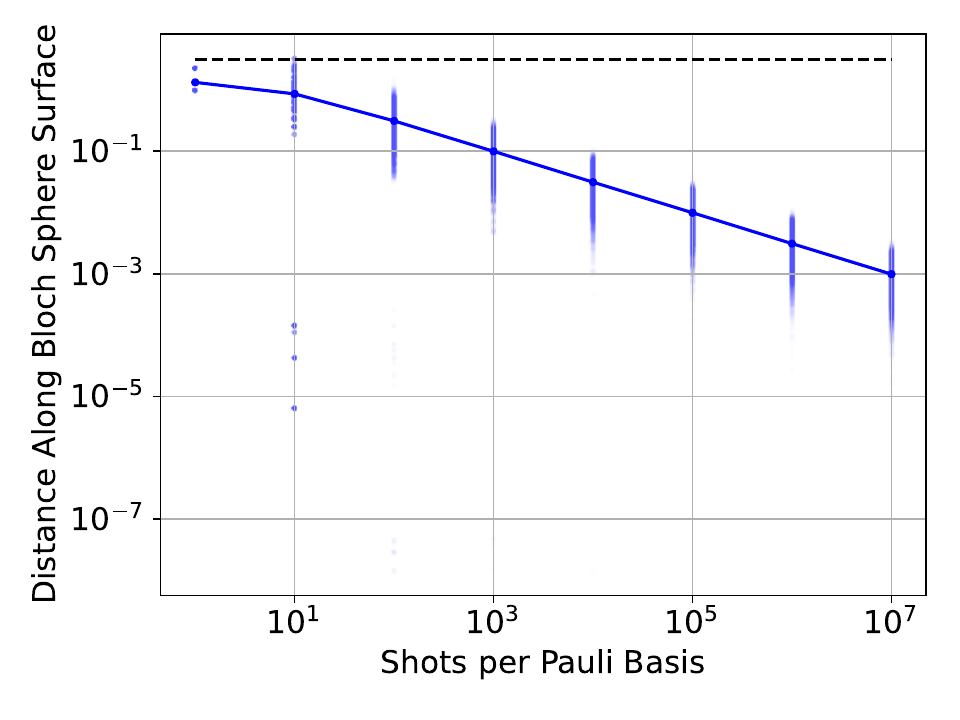}
    \includegraphics[width=0.49\textwidth]{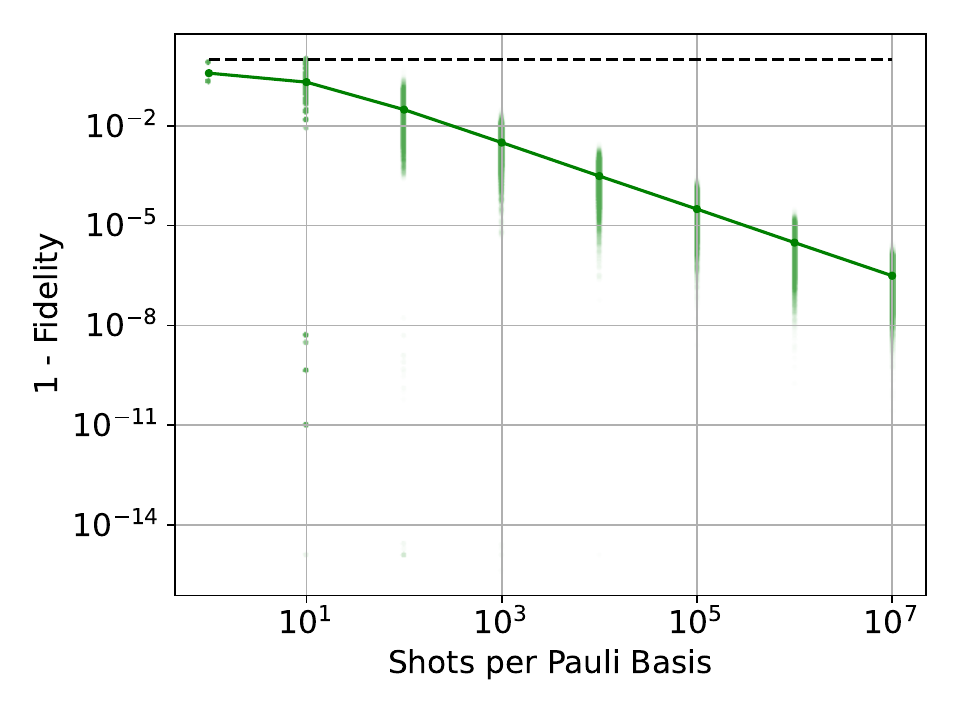}
    \caption{Full distribution of error rates (y-axis) for an $M=10$ quantum cloning distribution as the number of shots per Pauli basis is increased (x-axis). For each number of shots on the x-axis, using the clone emulation method $M \cdot 1000$ separate simulations are executed. Therefore, for each point that is plotted on the x-axis, a total of $M \cdot 1000$ error datapoints are gathered, thereby plotting the full distribution of error rates that result from the quantum cloning emulation and subsequent geometric extrapolation technique. The solid line connects the average error rate of the distributions at each shot count per Pauli basis (x-axis). The maximum possible error rate is shown by the horizontal dashed black line. Log-log scale axis. }
    \label{fig:distribution_of_error_rates}
\end{figure}

Figures~\ref{fig:compare_parallel_QST_to_emulate},~\ref{fig:compare_direct_QST_to_clone},~\ref{fig:increasing_M_error_rate} \ref{fig:breakeven} all present the mean error rates, however an important aspect of the scaling of this transmission protocol is the full distribution of error rates. Figure~\ref{fig:distribution_of_error_rates} shows what this distribution is for $M=10$ (using the clone emulation method), as the number of shots used in the quantum state tomography increases. 

\section{Example Quantum Cloning Circuit Diagram}
\label{section:appendix_quantum_telecloning_circuit}

\begin{figure}[h!]
    \centering
    \includegraphics[width=1.01\textwidth]{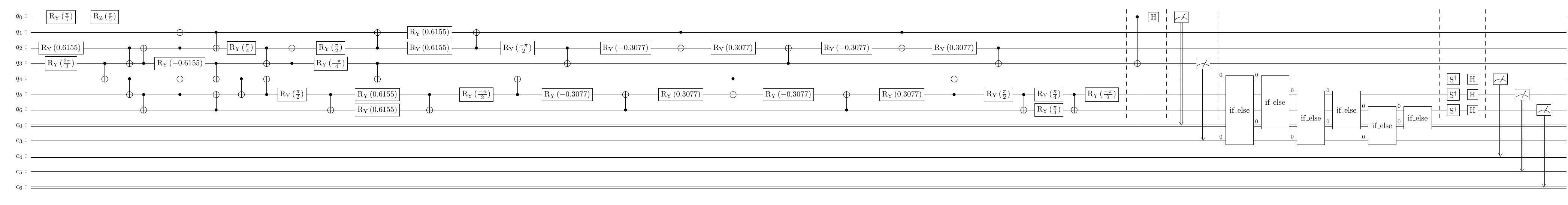}
    \caption{Circuit diagram from ref.~\cite{pelofske2023probing}. An example $1 \rightarrow 3$ quantum cloning circuit. This is specifically a quantum telecloning circuit thus requiring classical conditional and feed-forward operations, which implements ideal universal symmetric quantum cloning. Qubits 4, 5, 6 are the approximate copies of qubit 0. Qubit 3 is the port qubit, a required part of the quantum telecloning operation~\cite{pelofske2023probing, Pelofske_2022_clone1, Pelofske_2022_clone2}. Qubits 1 and 2 are ancilla qubits. The state that is prepared on qubit line 0 can be arbitrary - in this case it is two single qubit rotation gates to serve as a simple case. This particular circuit layout is targeting a linear line of qubits -- other qubit topologies could be used to implement the same unitary. The circuit barriers (vertical dashed lines) denote the logical phases of the quantum telecloning protocol, beginning with the quantum telecloning state preparation, followed by a Bell state measurement mid-circuit, followed by conditional if-else statements, followed by a rotation into the Pauli-Y basis (this basis change would rotate through different Pauli bases in order to perform full quantum state tomography), and ending with the measurement of the $3$ clone qubits.  }
    \label{fig:quantum_cloning_circuit}
\end{figure}

Figure~\ref{fig:quantum_cloning_circuit} shows an example quantum cloning circuit diagram for generating $3$ approximate copies of a single qubit. 

\clearpage

\setlength\bibitemsep{0pt}
\printbibliography

@article{PerezSalinas2020datareuploading,
  doi = {10.22331/q-2020-02-06-226},
  url = {https://doi.org/10.22331/q-2020-02-06-226},
  title = {Data re-uploading for a universal quantum classifier},
  author = {P{\'{e}}rez-Salinas, Adri{\'{a}}n and Cervera-Lierta, Alba and Gil-Fuster, Elies and Latorre, Jos{\'{e}} I.},
  journal = {{Quantum}},
  issn = {2521-327X},
  publisher = {{Verein zur F{\"{o}}rderung des Open Access Publizierens in den Quantenwissenschaften}},
  volume = {4},
  pages = {226},
  month = feb,
  year = {2020}
}

@misc{thumwanit2021trainable,
      title={Trainable Discrete Feature Embeddings for Variational Quantum Classifier}, 
      author={Napat Thumwanit and Chayaphol Lortaraprasert and Hiroshi Yano and Rudy Raymond},
      year={2021},
      eprint={2106.09415},
      archivePrefix={arXiv},
      primaryClass={quant-ph}
}

@misc{ambainis2009quantum,
      title={Quantum Random Access Codes with Shared Randomness}, 
      author={Andris Ambainis and Debbie Leung and Laura Mancinska and Maris Ozols},
      year={2009},
      eprint={0810.2937},
      archivePrefix={arXiv},
      primaryClass={quant-ph}
}

@article{Schmied_2016,
	doi = {10.1080/09500340.2016.1142018},
  
	url = {https://doi.org/10.1080%2F09500340.2016.1142018},
  
	year = 2016,
	month = {feb},
  
	publisher = {Informa {UK} Limited},
  
	volume = {63},
  
	number = {18},
  
	pages = {1744--1758},
  
	author = {Roman Schmied},
  
	title = {Quantum state tomography of a single qubit: comparison of methods},
  
	journal = {Journal of Modern Optics}
}

@misc{Qiskit,
    author = {{Qiskit contributors}},
    title = {Qiskit: An Open-source Framework for Quantum Computing},
    year = {2023},
    doi = {10.5281/zenodo.2573505}
}

@article{diamond2016cvxpy,
  author  = {Steven Diamond and Stephen Boyd},
  title   = {{CVXPY}: {A} {P}ython-embedded modeling language for convex optimization},
  journal = {Journal of Machine Learning Research},
  year    = {2016},
  volume  = {17},
  number  = {83},
  pages   = {1--5},
}

@article{agrawal2018rewriting,
  author  = {Agrawal, Akshay and Verschueren, Robin and Diamond, Steven and Boyd, Stephen},
  title   = {A rewriting system for convex optimization problems},
  journal = {Journal of Control and Decision},
  year    = {2018},
  volume  = {5},
  number  = {1},
  pages   = {42--60},
}

@article{agrawal2019dgp,
  author  = {Agrawal, Akshay and Diamond, Steven and Boyd, Stephen},
  title   = {Disciplined geometric programming},
  journal = {Optimization Letters},
  publisher = {Springer Berlin Heidelberg},
  year    = {2019},
  volume = {13},
  number = {5},
  pages = {961--976},
}

@article{agrawal2020dqcp,
    author       = {Agrawal, Akshay and Boyd, Stephen},
    title        = {Disciplined quasiconvex programming},
    journal      = {Optimization Letters},
    publisher = {Springer Berlin Heidelberg},
    year    = {2020},
}

@inproceedings{agrawal2019differentiable,
  title={Differentiable convex optimization layers},
  author={Agrawal, Akshay and Amos, Brandon and Barratt, Shane and Boyd, Stephen and Diamond, Steven and Kolter, J. Zico},
  booktitle={Advances in Neural Information Processing Systems},
  pages={9558--9570},
  year={2019},
}

@article{agrawal2020differentiating,
  title={Differentiating through log-log convex programs},
  author={Agrawal, Akshay and Boyd, Stephen},
  journal={arXiv},
  archivePrefix={arXiv},
  eprint={2004.12553},
  primaryClass={math.OC},
  year={2020},
}

@article{10.7717/peerj-cs.103,
     title = {SymPy: symbolic computing in Python},
     author = {Meurer, Aaron and Smith, Christopher P. and Paprocki, Mateusz and \v{C}ert\'{i}k, Ond\v{r}ej and Kirpichev, Sergey B. and Rocklin, Matthew and Kumar, AMiT and Ivanov, Sergiu and Moore, Jason K. and Singh, Sartaj and Rathnayake, Thilina and Vig, Sean and Granger, Brian E. and Muller, Richard P. and Bonazzi, Francesco and Gupta, Harsh and Vats, Shivam and Johansson, Fredrik and Pedregosa, Fabian and Curry, Matthew J. and Terrel, Andy R. and Rou\v{c}ka, \v{S}t\v{e}p\'{a}n and Saboo, Ashutosh and Fernando, Isuru and Kulal, Sumith and Cimrman, Robert and Scopatz, Anthony},
     year = 2017,
     month = jan,
     volume = 3,
     pages = {e103},
     journal = {PeerJ Computer Science},
     issn = {2376-5992},
     url = {https://doi.org/10.7717/peerj-cs.103},
     doi = {10.7717/peerj-cs.103}
    }

@article{pelofske2023probing,
   title={Probing Quantum Telecloning on Superconducting Quantum Processors},
   volume={5},
   ISSN={2689-1808},
   url={http://dx.doi.org/10.1109/TQE.2024.3391654},
   DOI={10.1109/tqe.2024.3391654},
   journal={IEEE Transactions on Quantum Engineering},
   publisher={Institute of Electrical and Electronics Engineers (IEEE)},
   author={Pelofske, Elijah and Bärtschi, Andreas and Eidenbenz, Stephan and Garcia, Bryan and Kiefer, Boris},
   year={2024},
   pages={1–19} }

@inproceedings{Pelofske_2022_clone1,
	doi = {10.1109/qce53715.2022.00083},
  
	url = {https://doi.org/10.1109%2Fqce53715.2022.00083},
  
	year = 2022,
	month = {sep},
  
	publisher = {{IEEE}
},
  
	author = {Elijah Pelofske and Andreas Bartschi and Bryan Garcia and Boris Kiefer and Stephan Eidenbenz},
  
	title = {Quantum Telecloning on {NISQ} Computers},
  
	booktitle = {2022 {IEEE} International Conference on Quantum Computing and Engineering ({QCE})}
}

@inproceedings{Pelofske_2022_clone2,
	doi = {10.1109/icrc57508.2022.00009},
  
	url = {https://doi.org/10.1109%2Ficrc57508.2022.00009},
  
	year = 2022,
	month = {dec},
  
	publisher = {{IEEE}
},
  
	author = {Elijah Pelofske and Andreas Bärtschi and Stephan Eidenbenz},
  
	title = {Optimized Telecloning Circuits: Theory and Practice of Nine {NISQ} Clones},
  
	booktitle = {2022 {IEEE} International Conference on Rebooting Computing ({ICRC})}
}

@misc{müller2023simplified,
      title={A Simplified Expression for Quantum Fidelity}, 
      author={Adrian Müller},
      year={2023},
      eprint={2309.10565},
      archivePrefix={arXiv},
      primaryClass={quant-ph}
}

@article{jozsa1994fidelity,
  title={Fidelity for mixed quantum states},
  author={Jozsa, Richard},
  journal={Journal of modern optics},
  volume={41},
  number={12},
  pages={2315--2323},
  year={1994},
  publisher={Taylor \& Francis}
}

@article{PhysRevLett.81.2598,
  title = {Optimal Universal Quantum Cloning and State Estimation},
  author = {Bruss, Dagmar and Ekert, Artur and Macchiavello, Chiara},
  journal = {Phys. Rev. Lett.},
  volume = {81},
  issue = {12},
  pages = {2598--2601},
  numpages = {0},
  year = {1998},
  month = {Sep},
  publisher = {American Physical Society},
  doi = {10.1103/PhysRevLett.81.2598},
  url = {https://link.aps.org/doi/10.1103/PhysRevLett.81.2598}
}

@article{dieks1982communication,
  title={Communication by EPR devices},
  author={Dieks, DGBJ},
  journal={Physics Letters A},
  volume={92},
  number={6},
  pages={271--272},
  year={1982},
  publisher={Elsevier},
  doi={10.1016/0375-9601(82)90084-6}
}

@article{wootters1982single,
  title={A single quantum cannot be cloned},
  author={Wootters, William K and Zurek, Wojciech H},
  journal={Nature},
  volume={299},
  pages={802--803},
  year={1982},
  publisher={Springer},
  doi={10.1038/299802a0}
}

@article{Bu_ek_1996,
	doi = {10.1103/physreva.54.1844},
  
	url = {https://doi.org/10.1103%2Fphysreva.54.1844},
  
	year = 1996,
	month = {sep},
  
	publisher = {American Physical Society ({APS})},
  
	volume = {54},
  
	number = {3},
  
	pages = {1844--1852},
  
	author = {V. Bu{\v{z}
}ek and M. Hillery},
  
	title = {Quantum copying: Beyond the no-cloning theorem},
  
	journal = {Physical Review A}
}

@article{Scarani_2005,
	doi = {10.1103/revmodphys.77.1225},
  
	url = {https://doi.org/10.1103%2Frevmodphys.77.1225},
  
	year = 2005,
	month = {nov},
  
	publisher = {American Physical Society ({APS})},
  
	volume = {77},
  
	number = {4},
  
	pages = {1225--1256},
  
	author = {Valerio Scarani and Sofyan Iblisdir and Nicolas Gisin and Antonio Ac{\'{\i}
}n},
  
	title = {Quantum cloning},
  
	journal = {Reviews of Modern Physics}
}

@article{PhysRevLett.79.2153,
  title = {Optimal Quantum Cloning Machines},
  author = {Gisin, N. and Massar, S.},
  journal = {Phys. Rev. Lett.},
  volume = {79},
  issue = {11},
  pages = {2153--2156},
  numpages = {0},
  year = {1997},
  month = {Sep},
  publisher = {American Physical Society},
  doi = {10.1103/PhysRevLett.79.2153},
  url = {https://link.aps.org/doi/10.1103/PhysRevLett.79.2153}
}

@article{fan2014quantum,
  title={Quantum cloning machines and the applications},
  author={Fan, Heng and Wang, Yi-Nan and Jing, Li and Yue, Jie-Dong and Shi, Han-Duo and Zhang, Yong-Liang and Mu, Liang-Zhu},
  journal={Physics Reports},
  volume={544},
  number={3},
  pages={241--322},
  year={2014},
  publisher={Elsevier}
}

@article{PhysRevA.67.022317,
  title = {Phase-covariant quantum cloning of qudits},
  author = {Fan, Heng and Imai, Hiroshi and Matsumoto, Keiji and Wang, Xiang-Bin},
  journal = {Phys. Rev. A},
  volume = {67},
  issue = {2},
  pages = {022317},
  numpages = {5},
  year = {2003},
  month = {Feb},
  publisher = {American Physical Society},
  doi = {10.1103/PhysRevA.67.022317},
  url = {https://link.aps.org/doi/10.1103/PhysRevA.67.022317}
}

@article{PhysRevA.62.012302,
  title = {Phase-covariant quantum cloning},
  author = {Bru\ss{}, Dagmar and Cinchetti, Mirko and Mauro D'Ariano, G. and Macchiavello, Chiara},
  journal = {Phys. Rev. A},
  volume = {62},
  issue = {1},
  pages = {012302},
  numpages = {7},
  year = {2000},
  month = {Jun},
  publisher = {American Physical Society},
  doi = {10.1103/PhysRevA.62.012302},
  url = {https://link.aps.org/doi/10.1103/PhysRevA.62.012302}
}

@article{PhysRevA.72.052322,
	doi = {10.1103/physreva.72.052322},
  
	url = {https://doi.org/10.1103%2Fphysreva.72.052322},
  
	year = 2005,
	month = {nov},
  
	publisher = {American Physical Society ({APS})},
  
	volume = {72},
  
	number = {5},
  
	author = {Thomas Durt and Jarom{\'{\i}
}r Fiur{\'{a}}{\v{s}}ek and Nicolas J. Cerf},
  
	title = {Economical quantum cloning in any dimension},
  
	journal = {Physical Review A}
}

@article{PhysRevLett.88.187901,
  title = {Approximate Quantum Cloning with Nuclear Magnetic Resonance},
  author = {Cummins, Holly K. and Jones, Claire and Furze, Alistair and Soffe, Nicholas F. and Mosca, Michele and Peach, Josephine M. and Jones, Jonathan A.},
  journal = {Phys. Rev. Lett.},
  volume = {88},
  issue = {18},
  pages = {187901},
  numpages = {4},
  year = {2002},
  month = {Apr},
  publisher = {American Physical Society},
  doi = {10.1103/PhysRevLett.88.187901},
  url = {https://link.aps.org/doi/10.1103/PhysRevLett.88.187901}
}

@article{PhysRevLett.86.4942,
	doi = {10.1103/physrevlett.86.4942},
  
	url = {https://doi.org/10.1103%2Fphysrevlett.86.4942},
  
	year = 2001,
	month = {may},
  
	publisher = {American Physical Society ({APS})},
  
	volume = {86},
  
	number = {21},
  
	pages = {4942--4945},
  
	author = {Jarom{\'{\i}
}r Fiur{\'{a}}{\v{s}}ek},
  
	title = {Optical Implementation of Continuous-Variable Quantum Cloning Machines},
  
	journal = {Physical Review Letters}
}

@misc{PhysRevLett.81.5003,
      title={Universal optimal cloning of qubits and quantum registers}, 
      author={Vladimir Buzek and Mark Hillery},
      year={1998},
      eprint={quant-ph/9801009},
      archivePrefix={arXiv},
      primaryClass={quant-ph},
  journal = {Phys. Rev. Lett.},
  volume = {81},
  issue = {22},
  pages = {5003--5006},
  numpages = {0},
  year = {1998},
  month = {Nov},
  publisher = {American Physical Society},
  doi = {10.1103/PhysRevLett.81.5003},
  url = {https://link.aps.org/doi/10.1103/PhysRevLett.81.5003}
}

@article{PhysRevLett.94.040505,
  title = {Experimental Quantum Cloning with Prior Partial Information},
  author = {Du, Jiangfeng and Durt, Thomas and Zou, Ping and Li, Hui and Kwek, L. C. and Lai, C. H. and Oh, C. H. and Ekert, Artur},
  journal = {Phys. Rev. Lett.},
  volume = {94},
  issue = {4},
  pages = {040505},
  numpages = {4},
  year = {2005},
  month = {Feb},
  publisher = {American Physical Society},
  doi = {10.1103/PhysRevLett.94.040505},
  url = {https://link.aps.org/doi/10.1103/PhysRevLett.94.040505}
}

@article{PhysRevA.72.042328,
  title = {Multipartite asymmetric quantum cloning},
  author = {Iblisdir, S. and Ac\'{\i}n, A. and Cerf, N. J. and Filip, R. and Fiur\'a\ifmmode \check{s}\else \v{s}\fi{}ek, J. and Gisin, N.},
  journal = {Phys. Rev. A},
  volume = {72},
  issue = {4},
  pages = {042328},
  numpages = {4},
  year = {2005},
  month = {Oct},
  publisher = {American Physical Society},
  doi = {10.1103/PhysRevA.72.042328},
  url = {https://link.aps.org/doi/10.1103/PhysRevA.72.042328}
}

@article{PhysRevA.60.136,
  title = {Strategies and networks for state-dependent quantum cloning},
  author = {Chefles, Anthony and Barnett, Stephen M.},
  journal = {Phys. Rev. A},
  volume = {60},
  issue = {1},
  pages = {136--144},
  numpages = {0},
  year = {1999},
  month = {Jul},
  publisher = {American Physical Society},
  doi = {10.1103/PhysRevA.60.136},
  url = {https://link.aps.org/doi/10.1103/PhysRevA.60.136}
}

@article{PhysRevLett.87.247901,
  title = {Telecloning of Continuous Quantum Variables},
  author = {van Loock, P. and Braunstein, Samuel L.},
  journal = {Phys. Rev. Lett.},
  volume = {87},
  issue = {24},
  pages = {247901},
  numpages = {4},
  year = {2001},
  month = {Nov},
  publisher = {American Physical Society},
  doi = {10.1103/PhysRevLett.87.247901},
  url = {https://link.aps.org/doi/10.1103/PhysRevLett.87.247901}
}

@article{hardy1999no,
  title={No signalling and probabilistic quantum cloning},
  author={Hardy, Lucien and Song, David D},
  journal={Physics Letters A},
  volume={259},
  number={5},
  pages={331--333},
  year={1999},
  publisher={Elsevier}
}

@article{PhysRevLett.106.180404,
  title = {Experimental Demonstration of Probabilistic Quantum Cloning},
  author = {Chen, Hongwei and Lu, Dawei and Chong, Bo and Qin, Gan and Zhou, Xianyi and Peng, Xinhua and Du, Jiangfeng},
  journal = {Phys. Rev. Lett.},
  volume = {106},
  issue = {18},
  pages = {180404},
  numpages = {4},
  year = {2011},
  month = {May},
  publisher = {American Physical Society},
  doi = {10.1103/PhysRevLett.106.180404},
  url = {https://link.aps.org/doi/10.1103/PhysRevLett.106.180404}
}

@article{PhysRevA.65.012304,
  title = {Quantum cloning machines for equatorial qubits},
  author = {Fan, Heng and Matsumoto, Keiji and Wang, Xiang-Bin and Wadati, Miki},
  journal = {Phys. Rev. A},
  volume = {65},
  issue = {1},
  pages = {012304},
  numpages = {7},
  year = {2001},
  month = {Dec},
  publisher = {American Physical Society},
  doi = {10.1103/PhysRevA.65.012304},
  url = {https://link.aps.org/doi/10.1103/PhysRevA.65.012304}
}

@article{PhysRevLett.126.060503,
  title = {All-Optical Optimal $N$-to-$M$ Quantum Cloning of Coherent States},
  author = {Liu, Shengshuai and Lou, Yanbo and Chen, Yingxuan and Jing, Jietai},
  journal = {Phys. Rev. Lett.},
  volume = {126},
  issue = {6},
  pages = {060503},
  numpages = {6},
  year = {2021},
  month = {Feb},
  publisher = {American Physical Society},
  doi = {10.1103/PhysRevLett.126.060503},
  url = {https://link.aps.org/doi/10.1103/PhysRevLett.126.060503}
}

@article{PhysRevA.59.156,
  title = {Quantum telecloning and multiparticle entanglement},
  author = {Murao, M. and Jonathan, D. and Plenio, M. B. and Vedral, V.},
  journal = {Phys. Rev. A},
  volume = {59},
  issue = {1},
  pages = {156--161},
  numpages = {0},
  year = {1999},
  month = {Jan},
  publisher = {American Physical Society},
  doi = {10.1103/PhysRevA.59.156},
  url = {https://link.aps.org/doi/10.1103/PhysRevA.59.156}
}

@article{bouchard2017high,
  title={High-dimensional quantum cloning and applications to quantum hacking},
  author={Bouchard, Fr{\'e}d{\'e}ric and Fickler, Robert and Boyd, Robert W and Karimi, Ebrahim},
  journal={Science advances},
  volume={3},
  number={2},
  pages={e1601915},
  year={2017},
  publisher={American Association for the Advancement of Science}
}

@article{PhysRevLett.105.073602,
  title = {Experimental Optimal Cloning of Four-Dimensional Quantum States of Photons},
  author = {Nagali, E. and Giovannini, D. and Marrucci, L. and Slussarenko, S. and Santamato, E. and Sciarrino, F.},
  journal = {Phys. Rev. Lett.},
  volume = {105},
  issue = {7},
  pages = {073602},
  numpages = {4},
  year = {2010},
  month = {Aug},
  publisher = {American Physical Society},
  doi = {10.1103/PhysRevLett.105.073602},
  url = {https://link.aps.org/doi/10.1103/PhysRevLett.105.073602}
}

@article{PhysRevA.66.052111,
  title = {Generation of phase-covariant quantum cloning},
  author = {Karimipour, V. and Rezakhani, A. T.},
  journal = {Phys. Rev. A},
  volume = {66},
  issue = {5},
  pages = {052111},
  numpages = {6},
  year = {2002},
  month = {Nov},
  publisher = {American Physical Society},
  doi = {10.1103/PhysRevA.66.052111},
  url = {https://link.aps.org/doi/10.1103/PhysRevA.66.052111}
}

@misc{fiurasek2005highly,
      title={Highly asymmetric quantum cloning in arbitrary dimension}, 
      author={Jaromir Fiurasek and Radim Filip and Nicolas J. Cerf},
      year={2005},
      eprint={quant-ph/0505212},
      archivePrefix={arXiv},
      primaryClass={quant-ph}
}

@article{Smolin_2012,
	doi = {10.1103/physrevlett.108.070502},
  
	url = {https://doi.org/10.1103%2Fphysrevlett.108.070502},
  
	year = 2012,
	month = {feb},
  
	publisher = {American Physical Society ({APS})},
  
	volume = {108},
  
	number = {7},
  
	author = {John A. Smolin and Jay M. Gambetta and Graeme Smith},
  
	title = {Efficient Method for Computing the Maximum-Likelihood Quantum State from Measurements with Additive Gaussian Noise},
  
	journal = {Physical Review Letters}
}

@article{PhysRevLett.119.170502,
  title = {Two-Hierarchy Entanglement Swapping for a Linear Optical Quantum Repeater},
  author = {Xu, Ping and Yong, Hai-Lin and Chen, Luo-Kan and Liu, Chang and Xiang, Tong and Yao, Xing-Can and Lu, He and Li, Zheng-Da and Liu, Nai-Le and Li, Li and Yang, Tao and Peng, Cheng-Zhi and Zhao, Bo and Chen, Yu-Ao and Pan, Jian-Wei},
  journal = {Phys. Rev. Lett.},
  volume = {119},
  issue = {17},
  pages = {170502},
  numpages = {6},
  year = {2017},
  month = {Oct},
  publisher = {American Physical Society},
  doi = {10.1103/PhysRevLett.119.170502},
  url = {https://link.aps.org/doi/10.1103/PhysRevLett.119.170502}
}

@article{PhysRevLett.101.080403,
  title = {Multistage Entanglement Swapping},
  author = {Goebel, Alexander M. and Wagenknecht, Claudia and Zhang, Qiang and Chen, Yu-Ao and Chen, Kai and Schmiedmayer, J\"org and Pan, Jian-Wei},
  journal = {Phys. Rev. Lett.},
  volume = {101},
  issue = {8},
  pages = {080403},
  numpages = {4},
  year = {2008},
  month = {Aug},
  publisher = {American Physical Society},
  doi = {10.1103/PhysRevLett.101.080403},
  url = {https://link.aps.org/doi/10.1103/PhysRevLett.101.080403}
}

@article{PhysRevA.90.032306,
  title = {Repeat-until-success quantum repeaters},
  author = {Bruschi, David Edward and Barlow, Thomas M. and Razavi, Mohsen and Beige, Almut},
  journal = {Phys. Rev. A},
  volume = {90},
  issue = {3},
  pages = {032306},
  numpages = {10},
  year = {2014},
  month = {Sep},
  publisher = {American Physical Society},
  doi = {10.1103/PhysRevA.90.032306},
  url = {https://link.aps.org/doi/10.1103/PhysRevA.90.032306}
}

@article{PhysRevLett.128.150502,
  title = {Optimal Entanglement Swapping in Quantum Repeaters},
  author = {Shchukin, Evgeny and van Loock, Peter},
  journal = {Phys. Rev. Lett.},
  volume = {128},
  issue = {15},
  pages = {150502},
  numpages = {5},
  year = {2022},
  month = {Apr},
  publisher = {American Physical Society},
  doi = {10.1103/PhysRevLett.128.150502},
  url = {https://link.aps.org/doi/10.1103/PhysRevLett.128.150502}
}

@article{PhysRevA.81.052329,
  title = {Efficient quantum repeater based on deterministic Rydberg gates},
  author = {Zhao, Bo and M\"uller, Markus and Hammerer, Klemens and Zoller, Peter},
  journal = {Phys. Rev. A},
  volume = {81},
  issue = {5},
  pages = {052329},
  numpages = {5},
  year = {2010},
  month = {May},
  publisher = {American Physical Society},
  doi = {10.1103/PhysRevA.81.052329},
  url = {https://link.aps.org/doi/10.1103/PhysRevA.81.052329}
}

@article{PhysRevA.77.062301,
  title = {Robust and efficient quantum repeaters with atomic ensembles and linear optics},
  author = {Sangouard, Nicolas and Simon, Christoph and Zhao, Bo and Chen, Yu-Ao and de Riedmatten, Hugues and Pan, Jian-Wei and Gisin, Nicolas},
  journal = {Phys. Rev. A},
  volume = {77},
  issue = {6},
  pages = {062301},
  numpages = {7},
  year = {2008},
  month = {Jun},
  publisher = {American Physical Society},
  doi = {10.1103/PhysRevA.77.062301},
  url = {https://link.aps.org/doi/10.1103/PhysRevA.77.062301}
}

@INPROCEEDINGS{9782866,

  author={Iqbal, Masab and Velasco, Luis and Ruiz, Marc and Napoli, Antonio and Pedro, Joao and Costa, Nelson},

  booktitle={2022 International Conference on Optical Network Design and Modeling (ONDM)}, 

  title={Quantum Bit Retransmission Using Universal Quantum Copying Machine}, 

  year={2022},

  volume={},

  number={},

  pages={1-3},

  doi={10.23919/ONDM54585.2022.9782866}}

@Article{s23187891,
AUTHOR = {Iqbal, Masab and Velasco, Luis and Costa, Nelson and Napoli, Antonio and Pedro, Joao and Ruiz, Marc},
TITLE = {Investigating Imperfect Cloning for Extending Quantum Communication Capabilities},
JOURNAL = {Sensors},
VOLUME = {23},
YEAR = {2023},
NUMBER = {18},
ARTICLE-NUMBER = {7891},
URL = {https://www.mdpi.com/1424-8220/23/18/7891},
PubMedID = {37765947},
ISSN = {1424-8220},
DOI = {10.3390/s23187891}
}

@article{wang2019duplicating,
  title={Duplicating classical bits with universal quantum cloning machine},
  author={Wang, MingHao and Cai, QingYu},
  journal={Science China Physics, Mechanics \& Astronomy},
  volume={62},
  pages={1--5},
  year={2019},
  publisher={Springer}
}

@misc{wang2018filling,
      title={Filling the gap between quantum no-cloning and classical duplication}, 
      author={Ming-hao Wang and Qing-yu Cai},
      year={2018},
      eprint={1803.05602},
      archivePrefix={arXiv},
      primaryClass={quant-ph}
}

@article{Pe_a_Tapia_2023,
	doi = {10.1088/1402-4896/acc5b8},
  
	url = {https://doi.org/10.1088%2F1402-4896%2Facc5b8},
  
	year = 2023,
	month = {apr},
  
	publisher = {{IOP} Publishing},
  
	volume = {98},
  
	number = {5},
  
	pages = {054001},
  
	author = {Elena Pe{\~{n}
}a Tapia and Giannicola Scarpa and Alejandro Pozas-Kerstjens},
  
	title = {A didactic approach to quantum machine learning with a single qubit},
  
	journal = {Physica Scripta}
}

@article{PhysRevLett.97.030402,
  title = {Asymptotic Quantum Cloning Is State Estimation},
  author = {Bae, Joonwoo and Ac\'{\i}n, Antonio},
  journal = {Phys. Rev. Lett.},
  volume = {97},
  issue = {3},
  pages = {030402},
  numpages = {4},
  year = {2006},
  month = {Jul},
  publisher = {American Physical Society},
  doi = {10.1103/PhysRevLett.97.030402},
  url = {https://link.aps.org/doi/10.1103/PhysRevLett.97.030402}
}

@book{paris2004quantum,
  title={Quantum state estimation},
  author={Paris, Matteo and Rehacek, Jaroslav},
  volume={649},
  year={2004},
  publisher={Springer Science \& Business Media}
}

@article{PhysRevA.57.2368,
  title = {Optimal universal and state-dependent quantum cloning},
  author = {Bru\ss{}, Dagmar and DiVincenzo, David P. and Ekert, Artur and Fuchs, Christopher A. and Macchiavello, Chiara and Smolin, John A.},
  journal = {Phys. Rev. A},
  volume = {57},
  issue = {4},
  pages = {2368--2378},
  numpages = {0},
  year = {1998},
  month = {Apr},
  publisher = {American Physical Society},
  doi = {10.1103/PhysRevA.57.2368},
  url = {https://link.aps.org/doi/10.1103/PhysRevA.57.2368}
}

@article{Bennett_2014,
   title={Quantum cryptography: Public key distribution and coin tossing},
   volume={560},
   ISSN={0304-3975},
   url={http://dx.doi.org/10.1016/j.tcs.2014.05.025},
   DOI={10.1016/j.tcs.2014.05.025},
   journal={Theoretical Computer Science},
   publisher={Elsevier BV},
   author={Bennett, Charles H. and Brassard, Gilles},
   year={2014},
   month=dec, pages={7–11} }

@article{Shor_2000,
   title={Simple Proof of Security of the BB84 Quantum Key Distribution Protocol},
   volume={85},
   ISSN={1079-7114},
   url={http://dx.doi.org/10.1103/PhysRevLett.85.441},
   DOI={10.1103/physrevlett.85.441},
   number={2},
   journal={Physical Review Letters},
   publisher={American Physical Society (APS)},
   author={Shor, Peter W. and Preskill, John},
   year={2000},
   month=jul, pages={441–444} }

@article{PhysRevA.72.012332,
  title = {Information-theoretic security proof for quantum-key-distribution protocols},
  author = {Renner, Renato and Gisin, Nicolas and Kraus, Barbara},
  journal = {Phys. Rev. A},
  volume = {72},
  issue = {1},
  pages = {012332},
  numpages = {17},
  year = {2005},
  month = {Jul},
  publisher = {American Physical Society},
  doi = {10.1103/PhysRevA.72.012332},
  url = {https://link.aps.org/doi/10.1103/PhysRevA.72.012332}
}

@misc{christandl2004genericsecurityproofquantum,
      title={A Generic Security Proof for Quantum Key Distribution}, 
      author={Matthias Christandl and Renato Renner and Artur Ekert},
      year={2004},
      eprint={quant-ph/0402131},
      archivePrefix={arXiv},
      primaryClass={quant-ph},
      url={https://arxiv.org/abs/quant-ph/0402131}, 
}

@inproceedings{https://doi.org/10.4230/lipics.tqc.2013.220,
  doi = {10.4230/LIPICS.TQC.2013.220},
  
  url = {https://drops.dagstuhl.de/entities/document/10.4230/LIPIcs.TQC.2013.220},
  
  author = {Yang, Yuxiang and Chiribella, Giulio},
  
  language = {en},
  
  title = {Is Global Asymptotic Cloning State Estimation?},
  
  publisher = {Schloss Dagstuhl – Leibniz-Zentrum für Informatik},
  
  year = {2013},
  
  copyright = {Creative Commons Attribution 3.0 Unported license}
}

\end{document}